\documentclass[11pt,twoside]{article}
\usepackage{asp2010}
\usepackage{lscape}
\usepackage{rotating}

\resetcounters

\begin{document}

\title{SD1000 Collaboration: Hunting down the subdwarf populations}

\author{Peter Nemeth, Roy {\O}stensen and Joris Vos}
\affil{Instituut voor Sterrenkunde, KU Leuven, Celestijnenlaan 200D,
3001 Leuven, Belgium; \texttt{pnemeth1981@gmail.com}}

\author{Adela Kawka and Stephane Vennes}
\affil{Astronomick\'{y} \'{u}stav AV \v{C}R, CZ-25165 Ond\v{r}ejov, Czech
Republic}

\begin{abstract}
Separating the subdwarf populations may shed light on their formation
history and give unique insights into the atmospheric processes that
are thought to be responsible for their abundance diversity. 
Such a task requires complex and time consuming spectral
analyses, which we started to tailor to our objectives a year ago. 
Here we report the updates that have been made over the past year
and will be the standard method in SD1000.
\end{abstract}

\section{Introduction}

Hot subdwarfs are core He burning stars at the blue end of the horizontal
branch (HB), also referred to as the extreme horizontal branch (EHB). 
These subluminous objects are located between the main-sequence (MS) and the
white dwarf (WD) sequence in the Hertzsprung-Russell diagram. 
The origin of hot subdwarfs is not clear, but an effective mass-loss
mechanism is certainly required to remove most of the envelope by the
time of core He ignition and leave behind an about half a solar mass star with a
very thin H envelope.
There is a heavy traffic of evolved stars in the region of subdwarfs in the
HRD and we see different populations of stars all at once. 
Some of these stars had different
initial masses and this increases the confusion regarding their 
population memberships.
Because of these problems and other limitations
caused by the low-number statistics that we currently experience, we
propose to initiate a large scale, statistically meaningful study of
subdwarf stars (SD1000).

\section{The pilot-study}

\begin{figure}[!h]
\begin{center}
\includegraphics[width=0.85\linewidth,bb=51 58 388 294 388 294,angle=0]{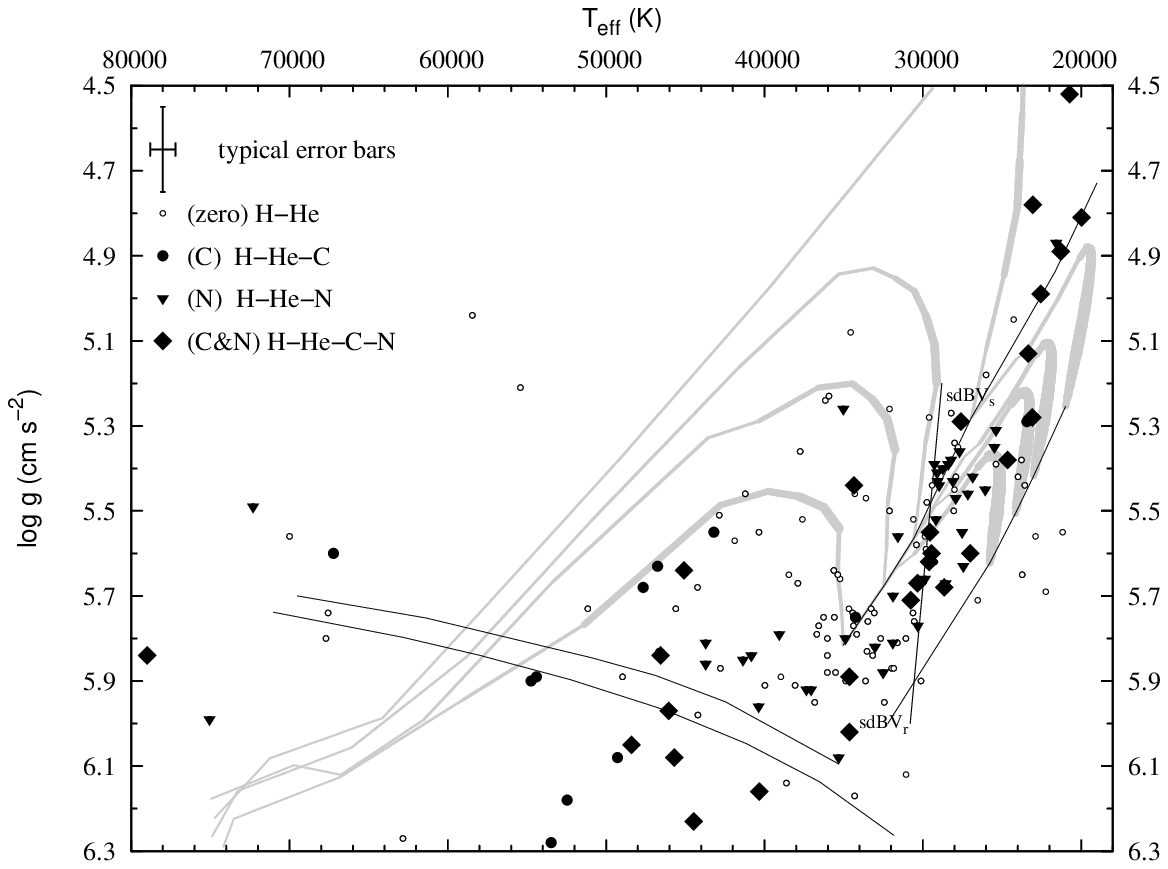}\\
\includegraphics[width=0.85\linewidth,bb=51 58 388 294 388 294,angle=0]{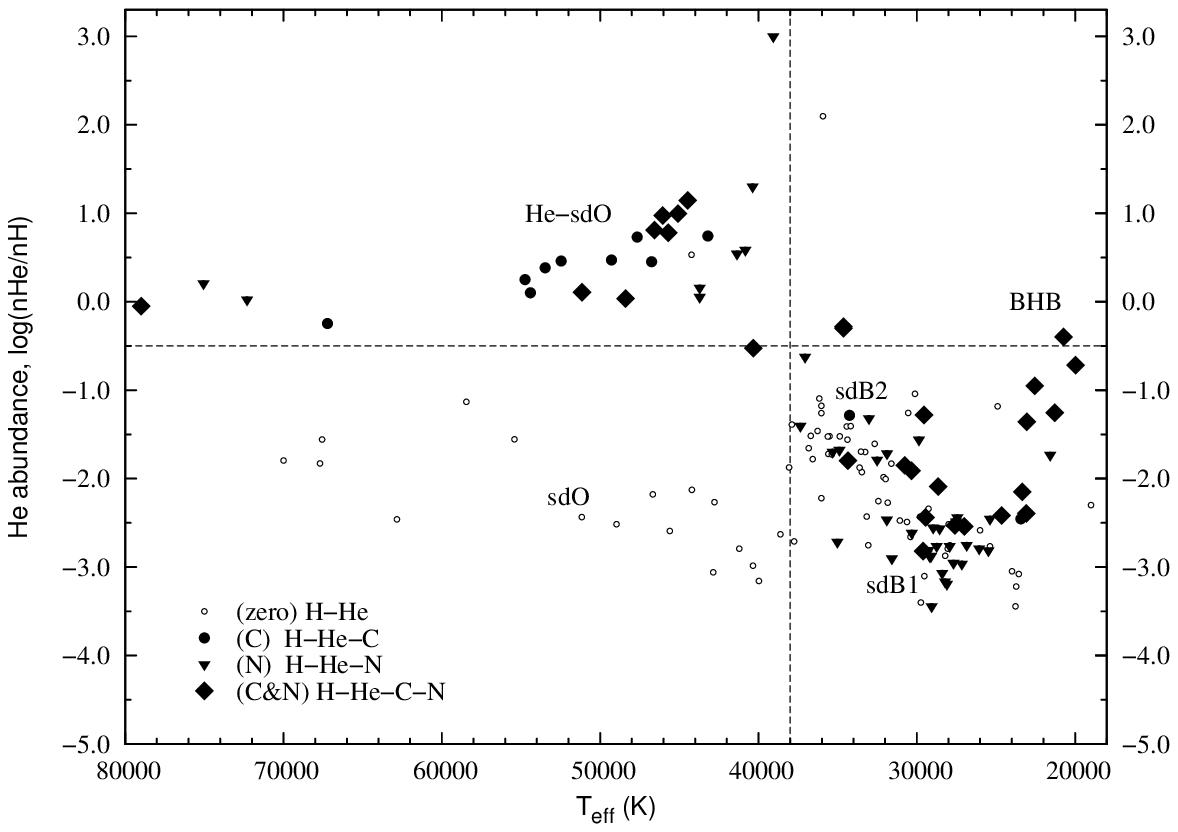}\\
\caption{ {\small
Metal abundances in the $T_{\rm eff}-\log g$ and $T_{\rm eff}-{\rm He}$
diagrams. }
\label{Fig:1} }
\end{center}   
\end{figure}   
\citet{2011MNRAS.410.2095V} initiated a programme to identify the white
dwarf population in the GALEX all-sky survey based on ultraviolet and
infrared color criteria. 
In their follow-up they collected and classified over 
200 high signal-to-noise (SNR) low-resolution
spectra of subluminous stars. 
The majority of these stars turned out to be bright hot subdwarfs. 
Later, \citet{nemeth12} developed a fitting procedure for {\sc Tlusty/Synspec}
(\citealt{hubeny95}; \citealt{lanz07}) which
included H/He/CNO opacities in the
subdwarf models and spectral decomposition for the composite spectra
binaries, and
carried out an NLTE model atmosphere analysis for 166 hot subdwarfs 
based on the follow-up spectroscopy.
The subsequent catalog is one of the largest homogeneously modeled sample to date. 
An intriguing result of this pilot-study was that the He abundance distribution
of hot subdwarfs 
shows a complex structure. 
It appears to be rather clumped in contrast to the previously observed 
simple linear trends.
More surprisingly sdB stars crowd in two groups that we define as: sdB1, near 
$T_{\rm eff}=28\,000$ K, $\log g=5.45$ dex, $\log(n{\rm He}/n{\rm H})=-2.7$; and
sdB2, near
$T_{\rm eff}=33\,500$ K, $\log g=5.80$ dex, $\log(n{\rm He}/n{\rm H})=-1.5$. 
Even more surprising was that these groups correlate with pulsational
properties and show a different CNO abundance pattern. 
It was found that sdB1 stars overlap with slow pulsators (sdBV${\rm _s}$)
and show N in their spectra. 
These stars are suspected close binaries from the common envelope 
formation channel and indeed, many short period binaries have been
previously found in this group. 
To further investigate this correlation 
a binary analysis for the GALEX sample is under way.
The sdB2 group overlaps with rapid pulsators (sdBV${\rm _r}$), 
does not show CNO
abundances and most of the composite spectra binaries were found among them. 
Such wide binaries may form through the Roche-lobe overflow channel. 
Hybrid, or mixed mode pulsators have been found mostly in 
between the sdB1 and sdB2 groups
where we see stars with substantial C and N abundances, which might be a
sign of diffusion or a different evolutionary channel. 
\citet{2013arXiv1307.5980G} found eight sdB stars with a He isotope anomaly
($^4$He is replaced by $^3$He)
in this region which also suggests the importance of element diffusion. 
Another group can be defined for to the blue horizontal branch (BHB) stars at 
$T_{\rm eff}<25\,000$ K, $\log g<5.2$ dex, $\log(n{\rm He}/n{\rm H})=-1$,
which also show C and N abundances. 
Similarly to sdB stars a complex structure was seen for He-sdO stars, 
for a reference please
see the abundance distributions in the $T_{\rm eff}-\log g$ and $T_{\rm
eff}-{\rm He}$ diagrams in Figure \ref{Fig:1} and \citet{nemeth12}.

This pilot-study proved that our method works and by applying for more
observations it can serve a better understanding of these subluminous stars.

\section{Abundance analysis}

Metal abundances play an important role in subdwarf spectroscopy. 
The He
abundances typical for sdB stars affect the NLTE atmosphere structure,
therefore the He abundance in sdB stars is a fundamental parameter and must be investigated
along with the surface temperature and gravity \citep{2009ARA&A..47..211H}.
To a lower extent CNO and Fe abundances
are also important in deriving accurate temperatures and gravities. 
Metal abundances also provide information on the formation of the stars
\citep{2012MNRAS.419..452Z}. 
\begin{sidewaysfigure}
\begin{center}
{\includegraphics[angle=0,width=0.9\textwidth]{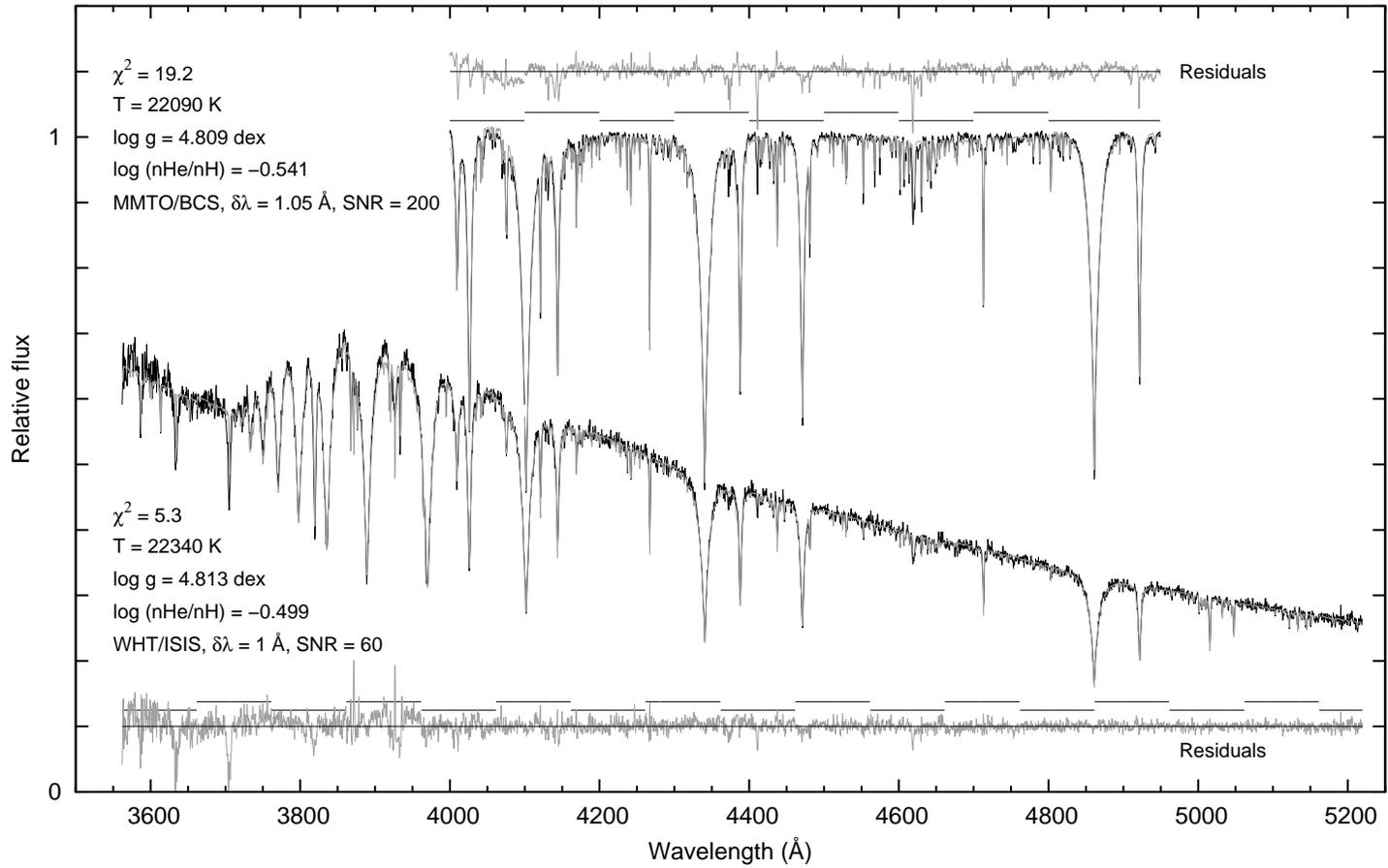}}\\
\caption{ {\small
Comparison of different high quality observations of the same star. 
The continuum normalized MMTO/BCS spectrum and the flux calibrated WHT/ISIS spectrum
were fitted separately and 
provide a consistent temperature, gravity and He abundance within
statistical errors.
The dashed lines show the regions where the synthetic spectra
(grey fit) were normalized to the observations (black). 
\label{Fig:2}}}
\end{center}
\end{sidewaysfigure}

While the CNO analysis in the GALEX subdwarf survey lead to important
conclusions, an extension to more elements, in particular including the 
iron group elements and metal line blanketing can open new horizons, 
but needs a more elaborated approach. 
Our new attempt uses atomic data for the
first 30 elements in the spectral synthesis and 14 elements in the
atmospheric structure calculations. 
With these many parameters the procedure is rather time consuming, but we 
could achieve a notable speed-up by
changing the opacity sampling parameters and the number of atmospheric
layers based on the goodness of fit. 
Figure \ref{Fig:2} shows a sample fit for a star at the tip of the 
blue horizontal branch. 
The MMTO/BCS (MMT Observatory / Blue Channel Spectrograph) data 
was obtained and kindly provided to us by Betsy Green (see also these
proceedings) and
will be discussed in detail in a forthcoming paper along with the WHT/ISIS (William
Herschel Telescope / Intermediate dispersion Spectrograph and Imaging System)
observation of the same star.

Element diffusion is necessary to explain the pulsations of sdB stars and
important in shaping the abundance profiles, hence 
diffusion calculations will certainly benefit from our 
homogeneous abundance determinations.

\section{Stellar rotation}

\begin{figure}[h]
\begin{center}
\includegraphics[width=\textwidth]{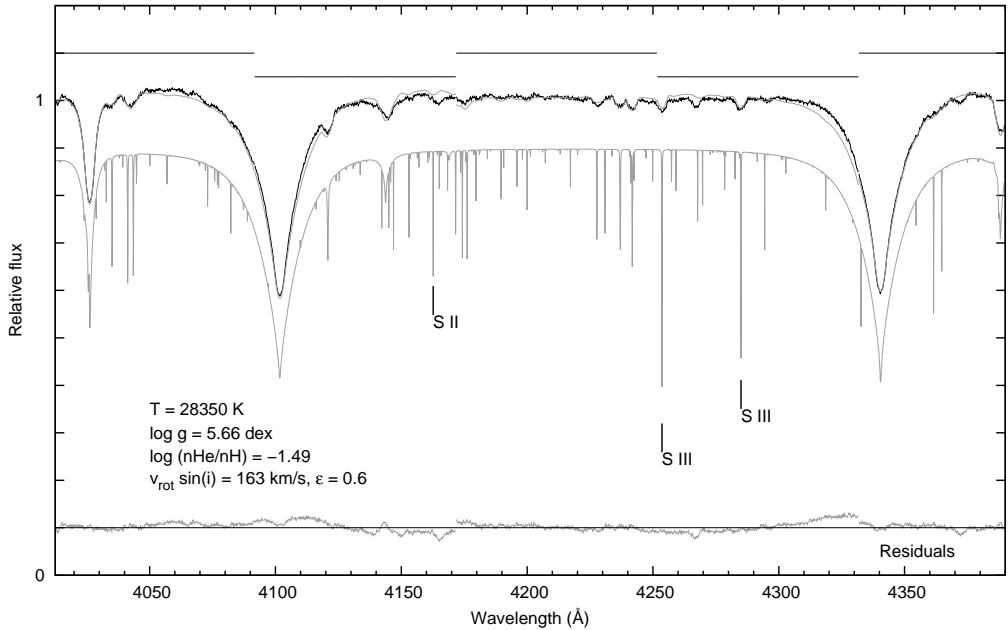}\\
\caption{ {\small
Our fit (grey) for the rapidly rotating sdB star
CD-3011223 (black), observed by \citet{2013A&A...554A..54G}. 
For comparison the emergent spectrum is also in this figure shifted
downwards. The strongest S lines are marked and the basic atmospheric
parameters are listed in the graph. The residual shows some mismatch in the
normalization of the co-added 174 individual spectra.
\label{Fig:3}}}
\end{center}
\end{figure}
Generally, a high rotational velocity is not typical for sdB stars. 
However, the number of exceptions increases and suggests 
that the common envelope evolution may not only lead to short-period
binaries (P$_{\rm orb}\sim0.1-10$ d), but to pre-cataclysmic binaries with
even shorter periods (P$_{\rm orb}<0.1$ d). 
In such systems tidal synchronization can result in a high rotational 
velocity of the primary. 
These binaries are promising Type Ia supernova progenitors and important to
understand.
In the GALEX subdwarf survey \citet{2012ApJ...759L..25V}
found that the radial velocity variation of CD-3011223 (GALEX J1411-3053)
is in excess of 300 km/s between two consecutive spectra taken over 20 minutes 
and the star showed rotationally broadened lines. 
Prompted by the discovery of the shortest period binary ever known we 
implemented rotational broadening
in our fitting procedure from \cite{gray08} and derived a $v_{\rm
rot}\sin{i}=165\pm5$ km/s, which is in good agreement with $v_{\rm  
rot}\sin{i}=177\pm10$ km/s independently found by \citet{2013A&A...554A..54G}. 
Figure \ref{Fig:3} shows our fit for the
SOAR/Goodman (Southern Astrophysical Research Telescope / Goodman
Spectrograph) spectrum.

\section{Systematics}

The sources of systematic biases are countless and even by only listing them
one could fill a paper. 
Finding the most important sources of these shifts is essential when working
with inhomogeneous data, therefore we will investigate such systematic
effects from a broader perspective in the future.
Here we are limited to mention only a few updates we already made. 

As can be noted in Figure \ref{Fig:2} our approach is to fit the entire
spectrum and seek for a global fit instead of some selected regions of the
data, because 
at a high resolution, high signal-to-noise and when metal line blanketing
is included each data point carries extra information and 
contribute to the fit. 
Hence, by consistently fitting the whole spectrum we are
able to track the interconnections between the H and the metal lines 
and reduce the Balmer line problem.
We found that a proper flux calibration and a larger spectral range 
also help the consistency of the derived parameters. 

The available atomic data we can work with is undoubtedly 
incomplete and suffer from inaccuracies. 
Line broadening data is often missing and not all lines can be taken with the
same profiles. 
A good example for this problem is the dependence of the microturbulent
velocity, which we consider an ad-hoc parameter, 
on the line formation depth or line strength. 
Another example is the instrumental broadening, which is usually taken with a Gaussian
profile, but a Gaussian is not necessarily the best representation.
Such broadenings affect the line profiles and the $\chi^2$, 
but not the equivalent widths of the lines. 
This means that a minimum $\chi^2$ might fit the observation well,
but still not represent the true surface parameters. 
We implemented a residual minimization in our procedure which can be used
independently or in a combination with a standard $\chi^2$ method. 
This approach minimizes the residual flux and is not so sensitive to spikes in the residual,
therefore some incorrect line profiles, like the He {\sc I} lines 
($\lambda3634$ and $\lambda3705$ \AA) in Figure \ref{Fig:2} 
cannot divert the procedure so easily.

Hot subdwarfs cover a wide range of temperatures and abundances, therefore
a grid calculated with a single set of atomic data may lead to an incorrect
ionization balance, rising systematic differences near the grid boundaries. 
An effective method to avoid this is to
calculate different grids for sdB, sdO and He-sdO stars, like in
\citet{2011MNRAS.410.2095V}.
Our original approach is a grid-less modeling, because calculating a well sampled grid with
iron group elements would be an endless undertaking, but
therefore the ionization equilibrium
must be continuously monitored and the atomic data input regularly adjusted to
the actual atmospheric conditions during fitting.  
Our experiences with this method are positive. 
The consistency of the
derived parameters between low and high-resolution, high SNR spectra are
better if the ionization balance is monitored. 
However, in some case a change in the level structure may lead to convergence problems.

Our latest update was to include the improved Stark
broadening tables of H lines from \cite{tremblay09} in the standard method. 

\section{SD1000 Collaboration}

The Collaboration is dedicated to process all the available
spectroscopic observations of hot subdwarfs to confirm the existing and set up
new constraints for theoretical work. 
Here we focus on optical spectroscopy, but where possible we wish to
complement the observations with ultraviolet data to derive more accurate
metal abundances.
As a result, we expect to better understand the contribution
of the various formation channels, the conditions of the mass loss on the
RGB and
be able to sort out extremely low mass WDs and post-AGB stars 
that are observed in the subdwarf region. 
Therefore the main objectives of the project are:  

\begin{itemize}
\item Spectroscopy for the largest possible sample for a better statistics and
distribution of hot subdwarfs. 
\item Perform a homogeneous analysis to find subtle trends and correlation among
surface parameters. Decrease instrumental biases.
\item Find connections to red giant stars and white dwarfs.
\item Provide accurate fundamental parameters for 
a statistically significant sample and with GAIA distances derive absolute
luminosities and masses. 
\item Find binary fractions and seek for 
connections between surface and binary parameters.
\end{itemize}

Currently we are working on some high-resolution, high signal-to-noise
spectra of sdB stars.
Interested observers can contribute to this target
sample and get updates at the project's web site:  
\begin{center}
\url{http://stelweb.asu.cas.cz/~nemeth/work/sd1000/}
\end{center}

\section{Conclusions}

To better understand the distribution of hot subdwarf stars in the
Hertzsprung-Russell diagram we
started a new collaborative programme (SD1000) and wish to process as many
high-quality spectroscopic observations as possible. 
By evaluating our efforts made in
the GALEX sample to derive accurate surface parameters we could 
improve our methods to derive more reliable effective temperatures, surface
gravities, abundances up to Z=30 as well as stellar rotation in a manageable way for
hundreds of stars. 

A promising direction is to apply our techniques to large spectroscopic datasets 
like the 
Bok-Green survey \citep{2008ASPC..392...75G}, 
SPY \citep{2001A&A...378L..17N}, 
MUCHFUSS (e.g: \citealt{2011A&A...530A..28G}; also these proceedings) and 
UVEX (e.g: \citealt{2012MNRAS.426.1235V}; also these proceedings) 
surveys as well as many individual, high quality observations. 
We also encourage others to perform such homogeneous analyses with independent methods for
reference.

Beyond minimizing the systematic shifts in subdwarf spectroscopy, 
future development will include the analysis of stratified atmospheres
to investigate element diffusion, which is a suspected
mechanism behind the observed abundance patterns. 
In order to resolve more MS companions of sdB stars and derive the radial velocities
of the components an important
contribution would be to extend the {\sc MILES} database 
\citep{2007MNRAS.374..664C} to the infrared and
include the Ca {\sc II} infrared triplet ($\lambda8498$, $\lambda8542$ and
$\lambda8662$ \AA). 
We will also consider synthetic spectra in this task. 

\acknowledgements{We thank our colleagues at the 
Armagh Observatory (Northern Ireland);
High Point University (NC, USA);
Dr. Remeis-Sternwarte, Bamberg (Germany);
European Southern Observatory, Garching (Germany) and at the
University of Arizona (AZ, USA)
for expressing their interest in the collaboration.
}

\end{document}